\documentclass[sigconf]{acmart}

\AtBeginDocument{%
  }

\usepackage{algorithm}
\usepackage{algorithmicx}
\usepackage{algpseudocode}
\usepackage{bm}
\usepackage{multirow}
\usepackage{xcolor}
\usepackage{epstopdf}
\usepackage[most]{tcolorbox}
\usepackage{tabularx}
\usepackage{enumitem}
\usepackage{hyperref}
\usepackage{balance}

\copyrightyear{2025}
\acmYear{2025}
\setcopyright{acmlicensed}\acmConference[KDD '25]{Proceedings of the 31st ACM SIGKDD Conference on Knowledge Discovery and Data Mining V.2}{August 3--7, 2025}{Toronto, ON, Canada}
\acmBooktitle{Proceedings of the 31st ACM SIGKDD Conference on Knowledge Discovery and Data Mining V.2 (KDD '25), August 3--7, 2025, Toronto, ON, Canada}
\acmDOI{10.1145/3711896.3737243}
\acmISBN{979-8-4007-1454-2/2025/08}

\begin{document}

\title{M$^2$-MFP: A Multi-Scale and Multi-Level Memory Failure Prediction Framework for Reliable Cloud Infrastructure}

\author{Hongyi Xie}
\authornote{This work was done when Hongyi Xie was an intern at Huawei.}
\affiliation{%
 \institution{University of Science and Technology of China}
 \city{Hefei}
 \country{China}}
\email{hyxie2023@mail.ustc.edu.cn}

\author{Min Zhou}
\authornote{Corresponding author.}
\affiliation{%
 \institution{Huawei}
 \city{Shenzhen}
 \country{China}}
\email{zhoumin27@huawei.com}

\author{Qiao Yu}
\affiliation{%
 \institution{Technical University of Berlin}
 \city{Berlin}
 \country{Germany}}
\email{qiao.yu@campus.tu-berlin.de}

\author{Jialiang Yu}
\affiliation{%
 \institution{Huawei}
 \city{Hangzhou}
 \country{China}}
\email{yujialiang@huawei.com}

\author{Zhenli Sheng}
\affiliation{%
 \institution{Huawei}
 \city{Hangzhou}
 \country{China}}
\email{shengzhenli@huawei.com}

\author{Hong Xie}
\authornotemark[2]
\affiliation{%
 \institution{University of Science and Technology of China}
 \city{Hefei}
 \country{China}}
\email{hongx87@ustc.edu.cn}

\author{Defu Lian}
\affiliation{%
 \institution{University of Science and Technology of China}
 \city{Hefei}
 \country{China}}
\email{liandefu@ustc.edu.cn}

\settopmatter{authorsperrow=4}

\begin{abstract}

As cloud services become increasingly integral to modern IT infrastructure, ensuring hardware reliability is essential to sustain high-quality service. Memory failures pose a significant threat to overall system stability, making accurate failure prediction through the analysis of memory error logs (i.e., Correctable Errors) imperative. Existing memory failure prediction approaches have notable limitations: rule-based expert models suffer from limited generalizability and low recall rates, while automated feature extraction methods exhibit suboptimal performance. To address these limitations, we propose M$^2$-MFP: a \textbf{M}ulti-scale and \textbf{M}ulti-Level \textbf{M}emory \textbf{F}ailure \textbf{P}rediction framework designed to enhance the reliability and availability of cloud infrastructure. M$^2$-MFP converts correctable errors (CEs) into multi-level binary matrix representations and introduces a Binary Spatial Feature Extractor (BSFE) to automatically extract high-order features at both DIMM-level and bit-level. Building upon the BSFE outputs, we develop a dual-path temporal modeling architecture: 1) a time-patch module that aggregates multi-level features within observation windows, and 2) a time-point module that employs interpretable rule-generation trees trained on bit-level patterns. Experiments on both benchmark datasets and real-world deployment show the superiority of M$^2$-MFP as it outperforms existing state-of-the-art methods by significant margins. Code and data are available at this repository: \href{https://github.com/hwcloud-RAS/M2-MFP}{https://github.com/hwcloud-RAS/M2-MFP}. 

\end{abstract}

\ccsdesc[300]{Computer systems organization~Reliability}
\ccsdesc[300]{Hardware~Failure prediction}

\keywords{Memory failure prediction; Event sequences; AIOps; Reliable cloud infrastructure}

\renewcommand{\shortauthors}{Hongyi Xie et al.}

\maketitle

\newcommand\kddavailabilityurl{https://doi.org/10.5281/zenodo.15515907}

\ifdefempty{\kddavailabilityurl}{}{
\begingroup\small\noindent\raggedright\textbf{KDD Availability Link:}\\
The source code of this paper has been made publicly available at \url{\kddavailabilityurl}.
\endgroup
}

\section{Introduction}

Modern cloud infrastructure depends critically on the sustained reliability of hardware components to provide uninterrupted services. With the exponential growth of data centers to meet escalating computational demands,
memory failures have become a predominant threat to overall system stability\cite{10.1145/3701716.3719148,hwang2012cosmic,sridharan2012study,wang2017can,notaro2023optical,wang2021workload}.
Unlike transient software issues, memory faults exhibit progressive error accumulation that existing error-correcting mechanisms cannot fully resolve. Analysis of Correctable Errors (CEs) reveals that their spatio-temporal patterns contain vital precursors to critical failures, yet current cloud systems predominantly employ reactive maintenance strategies\cite{schroeder2009dram,li2022correctable, liustim, yu2023himfp, criss2020improving}. 
This gap underscores the urgent need for proactive prediction frameworks to complement existing error correction capabilities.

Memory failure prediction aims to identify impending hardware faults by analyzing temporal and spatial patterns in memory error logs. This task involves modeling sequences of CEs, which are anomalies detected and corrected by error-checking mechanisms, to enable predictions before critical failures occur. The core challenges arise from three inherent complexities: 1) operational noise and data missingness in real-world error logs, 2) extreme class imbalance between normal operations and rare failure events, and 3) heterogeneous hardware configurations across manufacturers and architectures. These factors collectively require models capable of learning robust representations from sparse, noisy, and highly variable data sources. 

Traditional approaches have significant limitations in addressing these challenges. Rule-based expert systems\cite{du2018memory, du2020predicting, yu2023exploring} struggle to adapt to diverse hardware configurations and evolving failure patterns, requiring substantial expert involvement. Automated machine learning methods typically rely on hand-engineered features that do not capture spatial associations between error occurrences\cite{du2021predicting, boixaderas2020cost}. Existing deep learning models often treat error logs as continuous events without considering the hierarchical nature of memory architectures, leading to suboptimal feature representations\cite{liustim}. These methods demonstrate poor generalization across different memory types and manufacturers, limiting their practical deployment in heterogeneous cloud environments.

Although memory failure prediction has been extensively explored across cloud service providers and device manufacturers, the unavailability of a large-scale dataset that encompasses both physical (memory cell address) and logical (error bits during data access) information in error logs is primarily attributed to privacy concerns. To advance research in this field, we derive a benchmark dataset from Huawei Cloud's operational data centers. This dataset includes over 70,000 memory modules with CEs, of which more than 1,700 experienced failures. It spans a diverse range of CPU architectures, drawn from millions of servers over a nine-month period. In addition to the rich error logs information, the dataset also fully encapsulates real-world challenges like incomplete data, significant noise, class imbalance, and variability across manufacturers.



We further propose a \textbf{M}ulti-Scale and \textbf{M}ulti-Level \textbf{M}emory \textbf{F}ailure \textbf{P}rediction framework (M$^2$-MFP). The framework introduces a novel Multi-Level Binary Spatial Feature Extractor (Multi-BSFE) to derive representations of potential failure characteristics from CEs. Subsequently, M$^2$-MFP integrates a dual-path temporal modeling architecture comprising a \emph{Time-Patch Scale Prediction Module} for capturing local event patterns across temporal windows and a \emph{Time-Point Scale Prediction Module} for identifying crucial failure-inducing events at specific time points. This dual channel predictor demonstrates adaptability to predictions concerning both batch and streaming event sequence data. Compared to existing solutions, M$^2$-MFP exhibits enhanced intelligence, automation, and generalizability, necessitating minimal expert knowledge while significantly surpassing baselines in performance.


M$^2$-MFP has been officially deployed on Huawei Cloud's AIOps platform, continuously supporting the reliable operation and maintenance of over 400,000 servers. Our algorithm has demonstrated exceptional performance in Huawei Cloud's gray environment over three months, achieving an improvement in the $F_1$-score by approximately 15\%.

In summary, our main contributions are:
\begin{itemize}[leftmargin=*]
\item Publicly release the first large-scale memory logs dataset with multi-level CE information across diverse manufacturers and architectures, fostering widespread benefits and advancements in the field.
\item Propose M$^2$-MFP, a
generic and hierarchical failure prediction framework poised to offer universal, extendable, and scalable hardware failure prediction capabilities.
\item Deploy M$^2$-MFP in the live production environment to provide real-time monitoring services for millions of devices, enhancing the reliability and availability of cloud service and saving downtime costs.

\end{itemize}

\section{Background, Dataset and Task}

\subsection{Memory Failure Prediction System}

\begin{figure}[ht]
  \centering
  \includegraphics[width=0.9\linewidth]{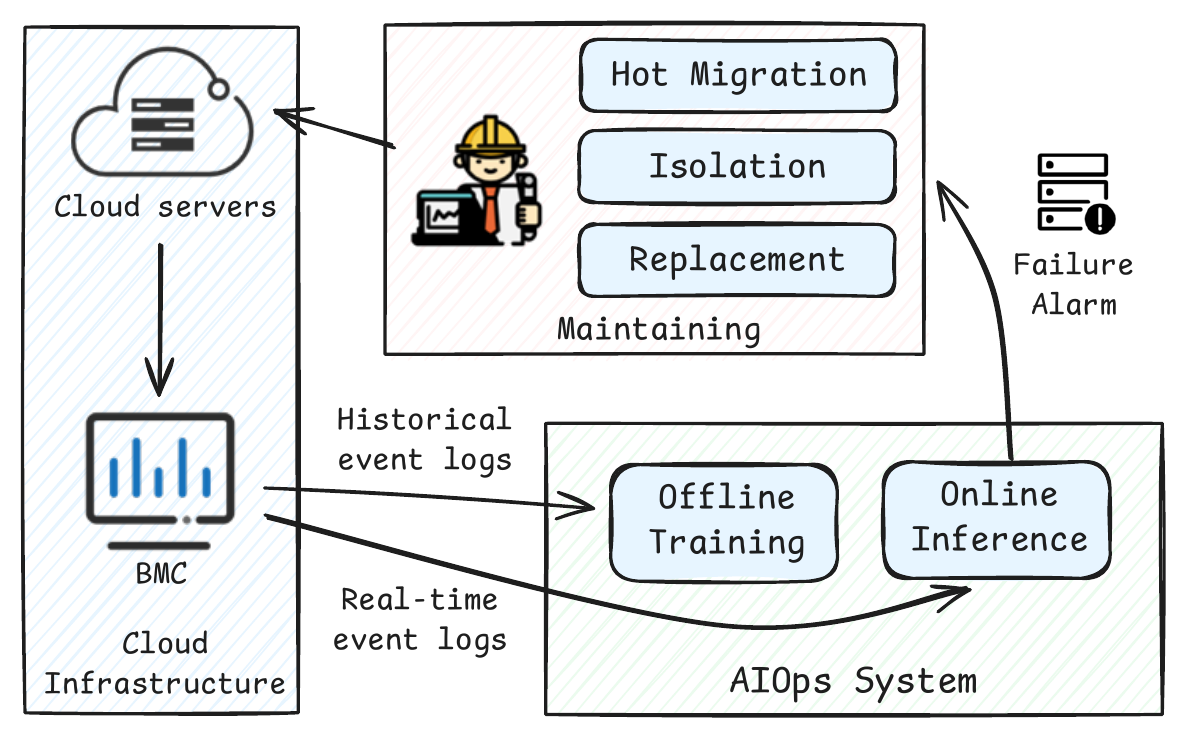}
  \caption{A Typical Failure Prediction Framework within the AIOps System}
  \Description{AIOps-System}
  \label{fig:AIOps-System}
\end{figure}

The memory failure prediction system forecasts future failures based on historical operational logs of cloud services, with Figure \ref{fig:AIOps-System} illustrating its workflow. The system relies on data from the Baseboard Management Controller (BMC), a dedicated microcontroller embedded in server motherboards that monitors and manages hardware, software, and network components of cloud services. For memory failure prediction, the BMC tracks memory operational status and transmits static configuration information (such as memory type, capacity, and manufacture) along with real-time event logs to the Artificial Intelligence for IT Operations (AIOps) system. Periodically (e.g., every three months in our deployment), we trains failure prediction models using offline-collected historical logs and deploys them online. These models then perform failure predictions at regular intervals (e.g., every 15 minutes) by analyzing event logs within sliding time windows. When predictions indicate potential failures, the system triggers alarms to notify architecture and hardware engineers for diagnostic analysis and corrective actions.

\begin{figure*}[ht]
  \centering
  \includegraphics[width=0.87\linewidth]{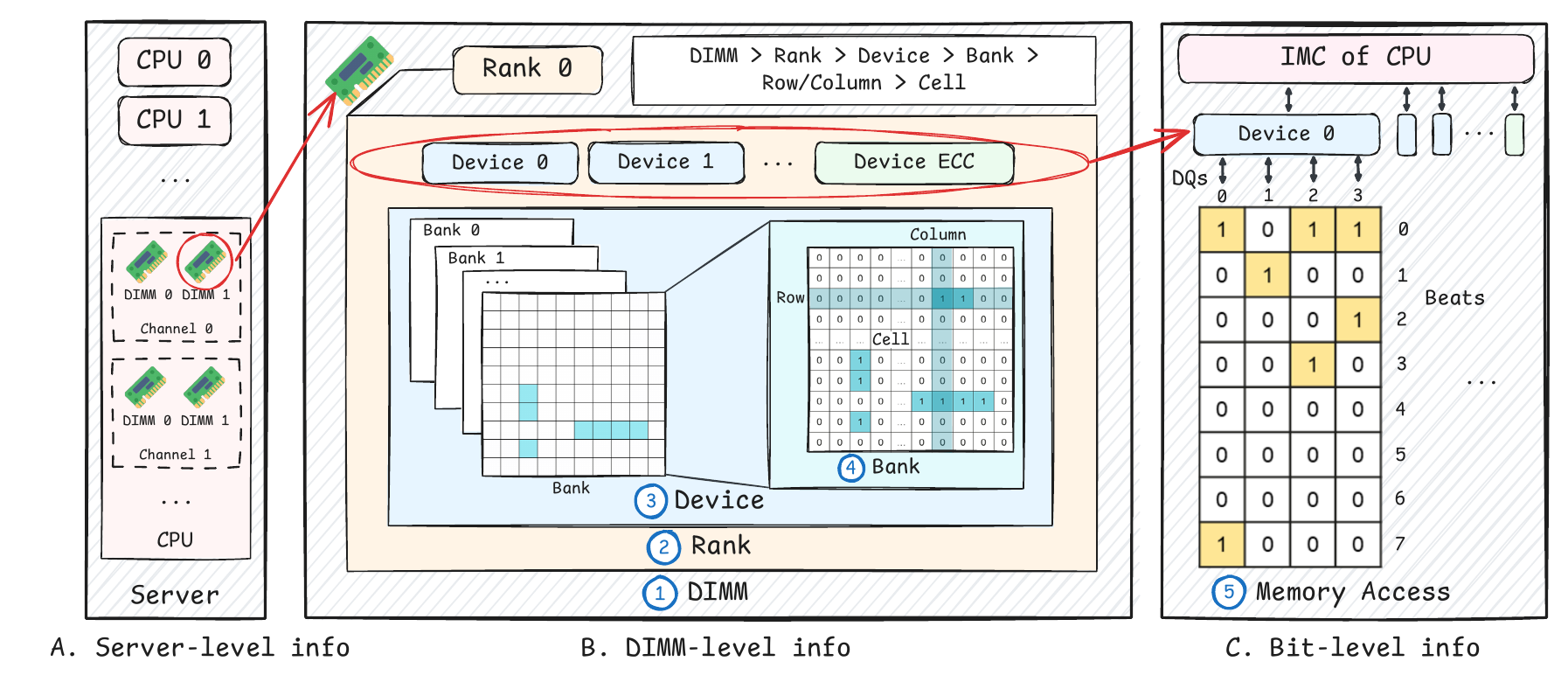}
  \caption{Hierarchical Levels Information of the Memory System}
  \Description{DRAM}
  \label{fig:DRAM}
\end{figure*}

Figure \ref{fig:DRAM} illustrates the hierarchical organization of memory systems in cloud servers, spanning  \textbf{Server-level} information (Fig. \ref{fig:DRAM}A), \textbf{DIMM-level} information (Fig. \ref{fig:DRAM}B), and \textbf{Bit-level} information (Fig. \ref{fig:DRAM}C). Each server integrates multiple CPUs, with server-level metadata recording CPU configurations and associated Dual In-line Memory Module (DIMM) modules. Fundamentally, a \large{\textcircled{\small{1}}}\normalsize DIMM comprises multiple DRAM chips (devices)\footnote{The terms "DRAM chips" and "devices" are used interchangeably in this paper.} grouped into \large{\textcircled{\small{2}}}\normalsize ranks to enable parallel read/write operations within the same rank. Each \large{\textcircled{\small{3}}}\normalsize device contains multiple parallel-operating \large{\textcircled{\small{4}}}\normalsize banks, which are further divided into rows and columns. The intersection of a row and a column forms a cell that stores a single data bit. The integrated memory controller (IMC) mediates communication between DIMM ranks and CPUs. For x4 DDR4 chips, each of the 4 DQ signal lines transmits 8 beats per access cycle, resulting in a total of 32 bits of data per transmission, which we represent as an $8 \times 4$ \large{\textcircled{\small{5}}}\normalsize DQ-Beat Matrix. Each rank includes an ECC chip for error detection/correction via ECC codes\cite{li2022correctable, rasguide}. While memory logs track correctable errors for predictive maintenance, catastrophic failures predominantly originate from uncorrectable transmission errors.

\subsection{Data Description}

Our dataset comprises logs from over 70,000 DIMMs in Huawei Cloud's production environments, collected between January and September 2024. The data includes two categories: CE logs and failure records of faulty DIMMs. An example of CE log is given below. Note that sensitive information, including manufacturer and region, has been anonymized in accordance with Huawei Cloud's confidentiality policy (see details in Appendix~\ref{sec:appendix-data}).

\begin{center}
\begin{tcolorbox}[colback=pink!20,
               colframe=black,
               width=8cm,	
               boxrule=1pt]
               \textcolor{black} 
		{\textbf{CE $e$ in DIMM $d$: }
		{[{'CpuId': 0, 'ChannelId': 1, 'DimmId': 0, 'RankId': 1, 'ChipId': 1, 'BankId': 2, 'RowId': 67745, 'ColumnId': 0,  'MemoryType': 'DDR4', 'Manufacturer': 'A', 'Region': 'E', 'Capacity': 16, 'ProcessorArchitecture': 'X86 Intel Purley', 'MaxSpeedMHz': 4000, 'FrequencyMHz': 2600, 'LogTime': 1711522709, 'beats': {'0': [], '1': [], '2': [], '3': [], '4': [], '5': [], '6': [57], '7': []}}]}}
\end{tcolorbox}
\end{center}

For a DIMM \(d\), let \(\mathcal{E}_d\) denote the collection of all CEs recorded for \(d\). Each CE \(e\) (where \(e \in \mathcal{E}_d\)) comprises an error type (either Read CE or Scrub CE), a log timestamp, and spatial information organized hierarchically into three levels:

\begin{itemize}[leftmargin=*]
    \item \textbf{Server-level:} Static configuration parameters of $d$, including server attributes such as CPU model, maximum frequency, base frequency, and DIMM-specific properties like DIMM model, capacity, and manufacturer specifications.

    \item \textbf{DIMM-level:} Identification of the failed memory cell, recording the fault location using hierarchical coordinates such as rank ID, device ID, bank ID, row ID, column ID.

    \item \textbf{Bit-level:} Error bits during memory access, captured as a 32-bit binary sequence representing the error state across the DQ-Beat Matrix for x4 DDR4 chips.  
\end{itemize}

\subsection{Memory Failure Prediction Task}

\begin{figure}[ht]
  \centering
  \includegraphics[width=0.9\linewidth]{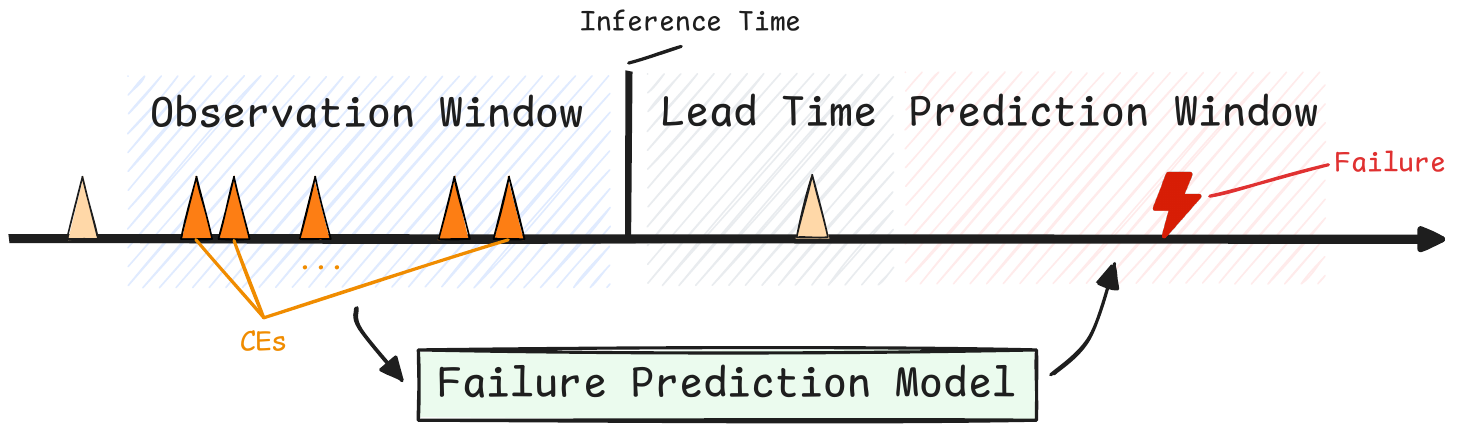}
  \caption{Memory Failure Prediction Task}
  \Description{MFP Task}
  \label{fig:MFP-Task}
\end{figure}

Memory fault prediction based on event sequences can be formulated as a binary classification task in machine learning. The goal is to predict whether a failure will occur within a specified future prediction window based on historical event information, as shown in Figure \ref{fig:MFP-Task}. The prediction incorporates a lead window, which is a critical buffer period accounting for real-world operational constraints, including data transmission delays, model inference latency, and the required response time for engineers to mitigate potential failures.

\noindent
\textbf{Problem formulation:} Given a DIMM \( d \) and its historical CE events \( \mathcal{E}_d \) (note that multiple CEs may occur simultaneously), the task is to dynamically predict, for any observation time \( t \), whether \( d \) will experience a failure within the future prediction window \([t + \Delta t_{\text{lead}}, t + \Delta t_{\text{lead}} + \Delta t_{\text{valid}}]\). Here, \( \Delta t_{\mathrm{lead}} \) represents the \emph{lead time}, which is the reserved period for taking preventive actions after predicting a failure, while \( \Delta t_{\mathrm{valid}} \) defines the \emph{validity window} for the prediction. The prediction utilizes all CEs in \( \mathcal{E}_d \) that occurred before \( t \), and the  binary target label \( y_d(t) \) is defined as \( y_d(t) = 1 \) if \( d \) fails within the interval, and \( y_d(t) = 0 \) otherwise.

\section{Related Works}

\subsection{Traditional Approaches}

Traditional memory failure prediction methods primarily rely on extracting features related to failure patterns based on expert experience or statistical analysis. \citet{schroeder2009dram} examined the correlation between static DRAM attributes and failures, and a series of subsequent works 78u6 \cite{hwang2012cosmic,meza2015revisiting,sridharan2012study,khan2016case,khan2016parbor,cheng2022depth,sridharan2015memory,beigi2023systematic,du2018memory,du2020predicting,li2007empirical} considered additional DIMM-level CE features. Based on analyses of memory mechanisms, \citet{giurgiu2017predicting} proposed the first memory failure prediction model using CEs, while \citet{boixaderas2020cost} built a machine learning model using CE count features. Other works \cite{yu2021dram,bogatinovski2022first,cheng2022depth} summarized DIMM-level spatial features of CE and employed statistical methods for feature construction combined with machine learning, and some works \cite{du2018memory,du2020predicting} explored more comprehensive DIMM-level CE features to establish rule-based models. However, these approaches are limited to DIMM-level CE features. \citet{li2022correctable} highlighted the importance of memory access in memory failure prediction by proposing the Risky CE pattern based on bit-level information, and \citet{yu2023exploring,yu2024icdcs,Yu2024investigating} further extended this work by designing additional memory access-related features and integrating both DIMM-level and bit-level CE features into the Himfp model \cite{yu2023himfp}. These methods, however, require substantial expert intervention or are heavily dependent on specific datasets, which may hinder their generalizability.

\subsection{Deep Learning Approaches}

Considering that the spatial features of CEs can be regarded as multi-level images, deep learning-based image processing methods such as Convolutional Neural Network (CNN) \cite{krizhevsky2012imagenet} and Vision Transformer (ViT) \cite{dosovitskiy2020image} can be employed to extract failure representations from CEs and train failure prediction models. STIM \cite{liustim} utilizes the Transformer \cite{vaswani2017attention} to effectively learn the hidden variability of bit-level CE features. This approach is not suitable for handling DIMM-level CE sparse spatial information, and it fails to accurately capture memory failure-related features, resulting in a high false positive rate. Moreover, while some deep learning-based failure prediction methods for hard disk drives \cite{hai2022hard,yang2020evaluating} are not well-suited for non-equidistant event sequences, generic event sequence failure prediction frameworks \cite{guo2021logbert,xu2019self,deng2024time,lu2024misp,lin2023edits,duan2024soil} struggle to mine the spatial information contained in memory failure logs; nevertheless, these methods have provided valuable inspiration.

\section{Methods}

Memory fault prediction methods typically comprise two stages: sample generation and classifier training. In the first stage, samples are generated from the raw CE event sequences by extracting features from both DIMM-level and bit-level information. The second stage involves training classifiers on these processed samples, where each sample is labeled as positive if it precedes a fault occurrence within a predefined future period, and negative otherwise. Based on the temporal characteristics of the CEs used for sample construction, the methods can be categorized into two groups: the \emph{time-patch} approach, which aggregates multiple CEs within a sliding time window, and the \emph{time-point} approach, which utilizes individual CE events for sample generation.

To effectively capture the spatial characteristics of CEs, we propose a \textbf{multi-level matrix representation} whose hierarchical structure and dimensional organization are detailed in Table~\ref{tab:fault_analysis}. This approach systematically organizes fault information across different abstraction levels. Each level in this hierarchical structure encapsulates fault patterns in a specific matrix format while maintaining explicit containment relationships with its subordinate levels, thereby preserving both the granular details and higher-level spatial correlations of error events.

\begin{table}[ht]
\captionsetup{skip=6pt}
\setlength{\tabcolsep}{3pt}
    \centering
    \caption{Fault Analysis Levels and Dimensions}
    \begin{tabular}{@{}lll@{}}
        \toprule
         & Level and Shape & Content \\ \midrule
        \multirow{4}{*}{DIMM-level} & DIMM ($1 \times N_\mathrm{rank}$) & Which ranks have faults \\
        & Rank ($1 \times N_\mathrm{device}$) & Which devices have faults \\
        & Device ($1 \times N_\mathrm{bank}$) & Which banks have faults \\
        & Bank ($N_\mathrm{row} \times N_\mathrm{col}$) & Which cells have faults \\ \midrule
        Bit-level & $N_\mathrm{beat} \times N_\mathrm{dq}$ matrix & Which bits have faults \\ \bottomrule
    \vspace{-0.5cm}
    \end{tabular}
    \label{tab:fault_analysis}
\end{table}


\subsection{M$^2$-MFP Framework}

\begin{figure*}[t]
  \centering
  \includegraphics[width=0.85\linewidth]{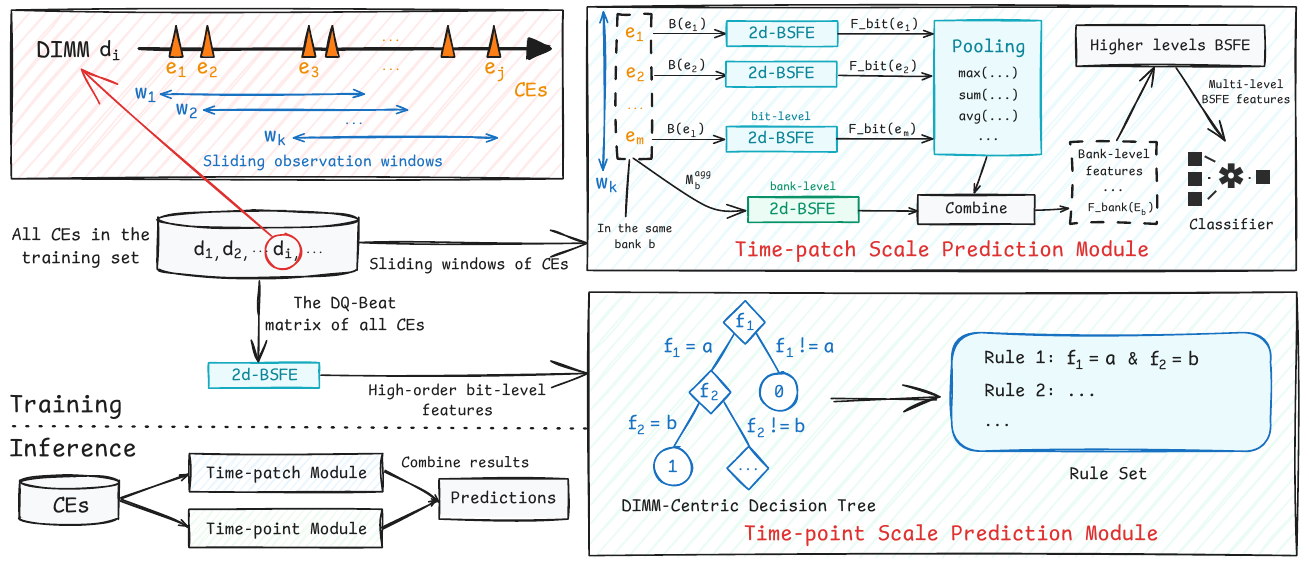}
  \caption{A Sketch of M$^2$-MFP Framework}
  \Description{M$^2$-MFP Framework}
  \label{fig:Framework}
\end{figure*}

The M$^2$-MFP framework introduces a novel dual-path architecture that combines multi-scale temporal analysis with multi-level feature representation, as illustrated in Figure~\ref{fig:Framework}. The framework begins with a Binary Spatial Feature Extractor (BSFE), designed to capture high-order features containing potential fault representations from the spatial information of CEs, as described in Subsection~\ref{sec:4.2}.

During the training stage, for time-patch scale data, we aggregate historical CEs using a sliding window and then apply multiple levels of BSFE through the \emph{Time-patch scale Prediction Module} to obtain multi-level spatial features (introduced in Subsection~\ref{sec:4.3}). For time-point scale data, we employ the \emph{Time-point scale Prediction Module} where BSFE first extracts bit-level high-order features from all CE data of each DIMM in the training set, after which a customized decision tree is trained to generate a rule set for failure prediction (introduced in Subsection~\ref{sec:4.4}).

During the inference stage, CE data are batch-processed through the Time-patch scale Prediction Module and streamed into the Time-point scale Prediction Module respectively. The final fault prediction results are obtained by merging outputs from both modules.

\subsection{Binary Spatial Feature Extractor (BSFE)}
\label{sec:4.2}

\begin{figure}[t]
  \centering
  \includegraphics[width=0.9\linewidth]{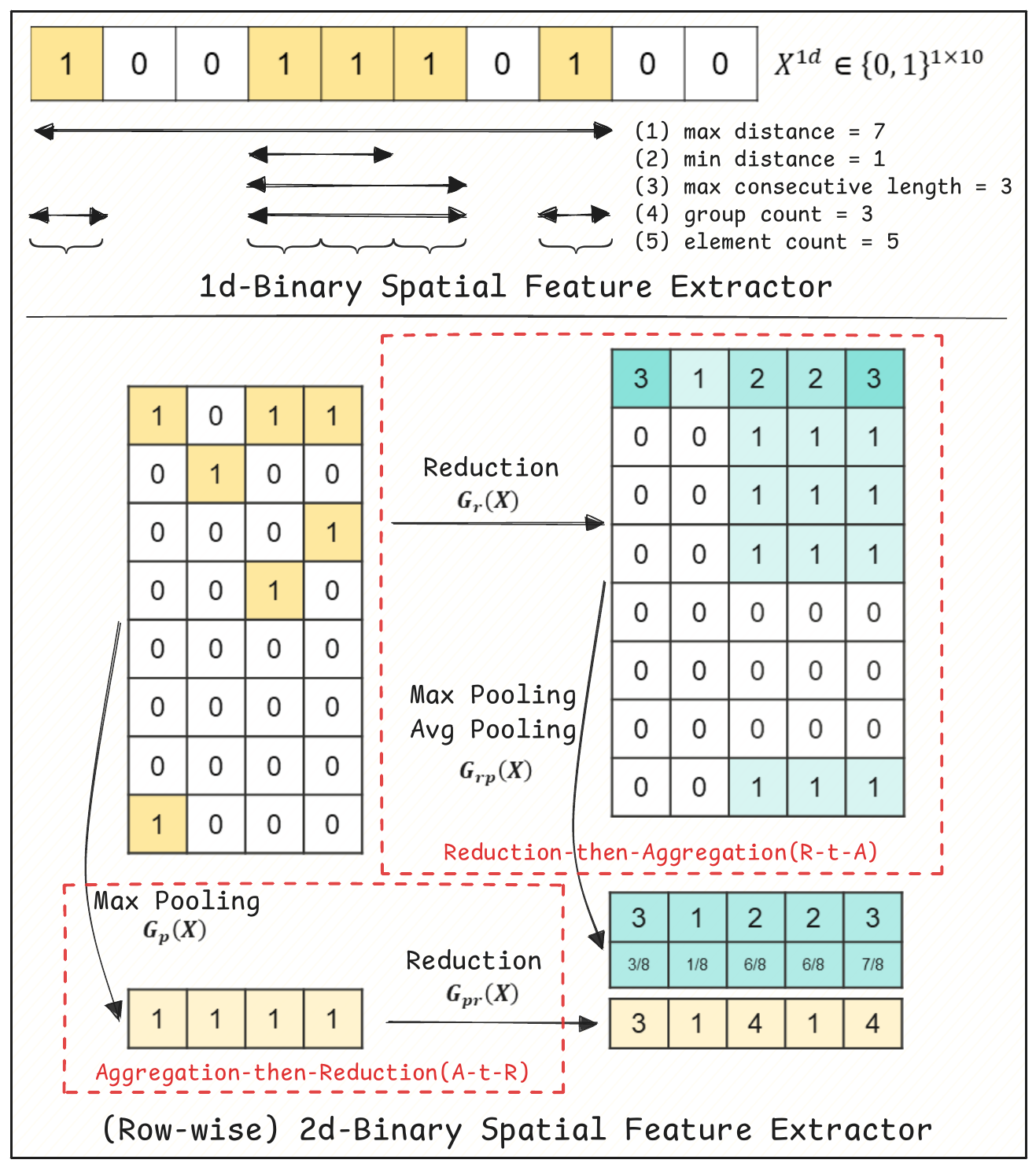}
  \caption{Binary Spatial Feature Extractor}
  \Description{BSFE}
  \vspace{-0.3cm}
  \label{fig:BSFE}
\end{figure}

We present a novel Binary Spatial Feature Extractor (BSFE) for extracting spatial information from CEs. The CE logs inherently contain spatial information at both the DIMM-level and bit-level, both of which can be effectively represented by binary matrices (0-1 matrices). To systematically capture this spatial information, we design BSFE, which consists of two components: one-dimensional BSFE (1d-BSFE) and two-dimensional BSFE (2d-BSFE), as illustrated in Figure \ref{fig:BSFE}. By leveraging custom functions and pooling, BSFE extracts meaningful spatial features from binary matrices. 
The design of BSFE is guided by the following principles:

\begin{itemize}[leftmargin=*]
    \item \textbf{Symmetry:} Given that the spatial distribution of CE logs exhibits symmetry, we ensure that the flipping of a one-dimensional matrix does not affect the results of 1d-BSFE.
    \item \textbf{Generality:} BSFE is designed to be applicable to binary matrices of varying dimensions and scales. It can be utilized at different granularity levels, including rank-level and device-level in one dimension, as well as bank-level and bit-level in two dimensions, ensuring adaptability across diverse spatial scales.
    \item \textbf{Sensitivity:} Considering that a single bit flip can impact the results of the ECC algorithm, we ensure that small perturbations in the binary matrix will reflect in the output of BSFE.
\end{itemize}

Given a binary matrix $X \in \{0, 1\}^{m \times n}$, we design BSFE features by considering symmetry, generality, and sensitivity. Specifically, we process each row or column of the matrix to extract spatial features that effectively capture element density, dispersion, clustering, and coverage, aiding in distinguishing fault distribution patterns in memory fault prediction. Formally, for a one-dimensional binary matrix 
\begin{equation}
    X^{1d} \triangleq [x_1, \ldots, x_n], \text{where}~x_i \in \{0,1\},
\end{equation}
the extracted features are computed as:
\begin{equation}
    BSFE^{1d}(X^{1d}) = 
    (\phi_1(X^{1d}), \ldots, \phi_f(X^{1d})),
\end{equation}
where $f$ represents the number of extracted features 
and $\phi_\ell (X^{1d}) \in \mathbb{R}$ denotes 
a feature descriptor.  
The following example illustrates five spatial descriptors: element count, group count, maximum consecutive count, maximum distance, and minimum distance.
\begin{align} 
    \textstyle\text{(element)}  & \textstyle\quad \phi_1(X^{1d}) = \sum_{i=1}^n x_i, \\
    \textstyle\text{(group)}    & \textstyle\quad \phi_2(X^{1d}) = \sum_{i=1}^{n-1} \mathbb{I}(x_i{=}0 \wedge x_{i+1}{=}1) + \mathbb{I}(x_1{=}1), \\
    \textstyle\text{(max-csc)} & \textstyle\quad \phi_3(X^{1d}) = \max_{1\leq i\leq j\leq n} (j-i+1)\prod_{k=i}^j x_k, \\
    \textstyle\text{(max-dist)} & \textstyle\quad \phi_4(X^{1d}) = \max_{1\leq i < j\leq n, x_i=x_j=1} |j-i|, \\
    \textstyle\text{(min-dist)} & \textstyle\quad \phi_5(X^{1d}) = \min_{1\leq i < j\leq n, x_i=x_j=1} |j-i|,
\end{align}
where $\mathbb{I}(\cdot)$ denotes the indicator function.


Building upon 1d-BSFE, we extend the approach to two-dimensional matrices to capture spatial dependencies at the bank and error bits. The 2d-BSFE consists of two modules: \textit{Reduction-then-Aggregation} and \textit{Aggregation-then-Reduction}. 

\begin{itemize}[leftmargin=*]
\item \textit{\textbf{Reduction-then-Aggregation}}:  
We first apply 1d-BSFE row-wise, yielding an intermediate representation:
\begin{equation}
    G_r(X) = \underset{\text{row-wise}}{BSFE}^{1d}(X), \quad G_r(X) \in \mathbb{R}^{m \times f}.
\end{equation}

Next, we apply a column-wise multiple pooling operation on \(G_r(X)\) using a pooling kernel of shape \(m \times 1\). Suppose we employ \(k\) different pooling methods (e.g., max pooling, average pooling, etc.); then the aggregated representation is
\begin{equation}
    G_{rp}(X) = \underset{\text{col-wise}}{Pool}^{m \times 1}(G_r(X)), \quad G_{rp}(X) \in \mathbb{R}^{k \times f}.
\end{equation}

\item 
\textit{\textbf{Aggregation-then-Reduction}}: 
We first apply a column-wise max pooling operation on the input \(X\) with a pooling kernel of shape \(m \times 1\) to obtain:
\begin{equation}
    G_p(X) = \underset{\text{col-wise}}{MaxPool}^{m \times 1}(X), \quad G_p(X) \in \mathbb{R}^{1 \times n}.
\end{equation}
Then, we apply 1d-BSFE row-wise to the max-pooled matrix:
\begin{equation}
    G_{pr}(X) = \underset{\text{row-wise}}{BSFE}^{1d}(G_p(X)), \quad G_{pr}(X) \in \mathbb{R}^{1 \times f}.
\end{equation}
\end{itemize}

The final representation of row-level 2d-BSFE, obtained by vectorizing \( G_{rp}(X) \) and \( G_{pr}(X) \) and then concatenating them, can be formally expressed as:
\begin{equation}
    \underset{\text{row-level}}{BSFE^{2d}(X)} = \left[ Vec(G_{rp}(X)), Vec(G_{pr}(X)) \right].
\end{equation}

By swapping row-wise and column-wise operations in the above procedure, we obtain the column-level 2d-BSFE. Finally, the overall 2d-BSFE is obtained by merging the row-level and column-level results:
\begin{equation}
    BSFE^{2d}(X) = \Bigl[ \underset{\text{row-level}}{BSFE^{2d}(X)},\, \underset{\text{column-level}}{BSFE^{2d}(X)} \Bigr].
\end{equation}

BSFE is essential for uncovering subtle patterns often overlooked by traditional rule-based methods, providing a solid foundation for memory fault analysis and prediction, thus enhancing the overall predictive capability of the framework.

\subsection{Time-Patch Scale Prediction Module}
\label{sec:4.3}

For the time-patch scale, we design a multi-level feature extractor called \textit{Multi-BSFE}, which combines hierarchical feature extraction with multi-scale aggregation via BSFE.
Given a set $\mathcal{E}$ of CE events within a time-patch, each event $e\in\mathcal{E}$ is first analyzed at the bit level. Let $B(e) \in \{0,1\}^{N_\mathrm{beat} \times N_\mathrm{dq}}$ denote the binary DQ-Beat Matrix of $e$. We extract bit-level features by
\begin{equation}
    F_{\mathrm{bit}}(e)=BSFE^{2d}\bigl(B(e)\bigr).
\end{equation}

At the bank level, events are grouped by their bank location. For a bank indexed by $b$, let $\mathcal{E}_b$ denote the set of all CEs in $\mathcal{E}$ whose Bank\_id equals $b$. Each CE records the position of a fault cell on a bank. This position is represented by a binary matrix $M_{\mathrm{b}} \in \{0,1\}^{N_{\mathrm{row}} \times N_{\mathrm{col}}}$, where exactly one element is 1, with its row and column indices indicating the fault cell's position. The aggregated bank matrix $M_b^{\mathrm{agg}}$ can be constructed via element-wise maximization:
\begin{equation}
    M_{\mathrm{b}}^{\mathrm{agg}}(r,c) {=} \max_{e \in \mathcal{E}_b} M_{\mathrm{b}}(e)(r,c), \quad \forall r {\in} \{1,\dots,N_{\mathrm{row}}\},\ c {\in} \{1,\dots,N_{\mathrm{col}}\}.
\end{equation}

The bank-level feature is then obtained by combining the BSFE of $M_b$ with the pooled (different pooling methods) high-order bit-level features:
\begin{equation}
    F_{\mathrm{bank}}(\mathcal{E}_b) = \left[
    Vec\Bigl( BSFE^{2d}\bigl(M_{\mathrm{b}}^{\mathrm{agg}}\bigr) \Bigr),\\
    Vec\Bigl(Pool\big(\{F_{\mathrm{bit}}(e)\}_{e\in\mathcal{E}_b}\big)\Bigr) \right].
\end{equation}

This hierarchical procedure is extended across the memory organization levels. For any two adjacent levels, a lower level $L$ and a higher level $H$ (e.g., bit-level $<$ Bank $<$ Device $<$ Rank $<$ DIMM), Multi-BSFE is applied as
\begin{equation}
    F_\mathrm{H}(\mathcal{E})= \left[Vec\Bigl(BSFE^{2d}\bigl(M_\mathrm{H}^{\mathrm{agg}}\bigr)\Bigr),\\
    Vec\Bigl(Pool\Bigl(\{F_\mathrm{L}(e)\}_{e\in\mathcal{E}}\Bigr)\Bigr) \right].
\end{equation}
where $M_\mathrm{H}^{\mathrm{agg}}$ is the aggregation matrix at level $H$.

In practice, since over 90\% of DIMMs exhibit only a single erroneous Device, Rank, or DIMM within a time-patch, we omit pooling for these low-dimensional levels and apply pooling only at the DIMM level.  
We also retain traditional counting features \cite{yu2023himfp, li2022correctable}, including:
\begin{itemize}
    \item CE counts within the time-patch,
    \item The count of events with DQ errors, denoted $\theta_{\mathrm{DQ}}$,
    \item The count of events with Beat errors, denoted $\theta_{\mathrm{Beat}}$,
    \item CE frequency (number of CE events per time unit).
\end{itemize}

The final time-patch feature vector is formed by concatenating the outputs of the Multi-BSFE with the derived counting features. These generated features are then fed into a classifier, such as LightGBM\cite{ke2017lightgbm}, to train a failure prediction model. During training and inference, features are generated at regular intervals of $\Delta i_p$ using CE logs from the preceding observation window, thereby ensuring temporal consistency in predictions.

\subsection{Time-Point Scale Prediction Module} 
\label{sec:4.4}

We designed a customized Gini decision tree based on bit-space features and CE\_type to achieve fault prediction at the time-point granularity. The key innovation lies in the utilization of DIMM-Centric Gini gain rather than sample-level impurity metrics. The algorithm is detailed below:

\begin{algorithm}[ht]
\caption{DIMM-Centric Decision Tree}
\begin{algorithmic}[1]
\Require Training data $D$ with DIMM IDs and features $F$
\Ensure Decision tree for time-point predictions

\State \textbf{DIMM Gini:}
\[
\mathcal{G}(\mathcal{D}) = 1 - \left(\frac{|\mathcal{D}^+|}{|\mathcal{D}|}\right)^2 - \left(\frac{|\mathcal{D}^-|}{|\mathcal{D}|}\right)^2
\]
where $\mathcal{D}$: DIMM set, $\mathcal{D}^+$/$\mathcal{D}^-$: faulty/normal DIMMs

\Procedure{BuildTree}{$D, F$}
    \State $\mathcal{D} \gets$ unique DIMMs in $D$, $\mathcal{D}^+ \gets$ faulty DIMMs
    \If{($\frac{|\mathcal{D}^+|}{|\mathcal{D}^-|} > \theta$ or $\frac{|\mathcal{D}^-|}{|\mathcal{D}^+|} > \theta$) or (Max depth reached or $|\mathcal{D}| \leq 1$)}
        \State \Return Leaf: $\mathbb{I}(|\mathcal{D}^+| \geq |\mathcal{D}^-|)$
    \EndIf
    
    \State Find optimal split $(f^*, v^*)$ minimizing:
    \[
    \min_{f,v} \left[ \frac{|\mathcal{D}_L|}{|\mathcal{D}|}\mathcal{G}(\mathcal{D}_L) + \frac{|\mathcal{D}_R|}{|\mathcal{D}|}\mathcal{G}(\mathcal{D}_R) \right]
    \]
    \State where $\mathcal{D}_L = \{d \in \mathcal{D} \mid \exists x \in d: x_f = v\}$, $\mathcal{D}_R = \mathcal{D} \setminus \mathcal{D}_L$
    
    \If{No valid split} 
        \Return Leaf
    \EndIf
    \State Partition $D$ into $D_L$ (DIMMs $\in \mathcal{D}_L$) and $D_R$
    \State \Return Node($f^*, v^*$, BuildTree($D_L$), BuildTree($D_R$))
\EndProcedure
\end{algorithmic}
\end{algorithm}

We extract all branches of the DIMM-Centric Decision Tree that yield a result of 1 (indicating a fault) into a rule base. During inference, each incoming CE data record from the data stream is evaluated against the rule base. If a record matches any rule, the corresponding DIMM is predicted to experience a fault in the future.

\section{Experiments}

\begin{table*}[ht]
\setlength{\tabcolsep}{3pt}
\centering
\caption{Performance comparison among different memory failure prediction models}
\label{tab:results_all}
\begin{tabular}{@{}ll*{9}{c}@{}}
\toprule
\multirow{2}{*}{Category} & \multirow{2}{*}{Method} & \multicolumn{3}{c}{Intel Purley} & \multicolumn{3}{c}{Intel Whitley} & \multicolumn{3}{c}{all} \\
\cmidrule(lr){3-5} \cmidrule(lr){6-8} \cmidrule(lr){9-11}
 &  & Precision & Recall & $F_1$-score & Precision & Recall & $F_1$-score & Precision & Recall & $F_1$-score \\
\midrule
\multirow{5}{*}{Time-point} 
& Naive               & 0.2085 & 0.2444 & 0.2250 & 0.0698 & 0.0417 & 0.0522 & 0.2024 & 0.2269 & 0.2140 \\
& Risky CE            & 0.0476 & 0.4875 & 0.0868 & 0.1088 & 0.3750 & 0.1686 & 0.0494 & 0.4778 & 0.0895 \\
& DQ Beat Predictor   & 0.0487 & 0.5059 & 0.0888 & 0.1020 & 0.3750 & 0.1603 & 0.0503 & 0.4946 & 0.0913 \\
& CNN                 & 0.0795 & 0.2168 & 0.1164 & 0.1681 & 0.2917 & \color{cyan}\underline{0.2133} & 0.0836 & 0.2185 & 0.1209 \\
& Time-point Ours     & 0.4047 & 0.2378 & \color{cyan}\underline{0.2996} & 0.3529 & 0.0833 & 0.1348 & 0.4029 & 0.2245 & \color{cyan}\underline{0.2883} \\
\midrule
\multirow{7}{*}{Time-patch}
& CNN                 & 0.0796 & 0.2838 & 0.1244 & 0.1053 & 0.2778 & 0.1527 & 0.0811 & 0.2689 & 0.1246 \\
& CNN (1D kernal)     & 0.0856 & 0.2799 & 0.1311 & 0.1563 & 0.1389 & 0.1471 & 0.0870 & 0.2665 & 0.1311 \\
& ViT                 & 0.0596 & 0.0302 & 0.0401 & 0.2308 & 0.0556 & 0.0896 & 0.0516 & 0.0360 & 0.0424 \\
& ViT (1D patch)      & 0.0759 & 0.3798 & 0.1265 & 0.1197 & 0.1944 & 0.1481 & 0.0771 & 0.3649 & 0.1273 \\
& STIM                & 0.0546 & 0.0644 & 0.0591 & 0.0641 & 0.1389 & 0.0877 & 0.0560 & 0.0708 & 0.0626 \\
& Himfp               & 0.1790 & 0.3246 & 0.2307 & 0.1918 & 0.1944 & 0.1931 & 0.1806 & 0.3097 & 0.2282 \\
& Time-patch Ours     & 0.3436 & 0.3022 & {\color{blue}\underline{0.3216}} & 0.2542 & 0.2222 & {\color{red}\textbf{0.2372}} & 0.3446 & 0.2893 & {\color{blue}\underline{0.3145}} \\
\midrule
\multirow{1}{*}{Combined}
& M$^2$-MFP       & 0.3344 & 0.3942 & {\color{red}\textbf{0.3619}} & 0.2500 & 0.2222 & {\color{blue}\underline{0.2353}} & 0.3208 & 0.3942 & {\color{red}\textbf{0.3537}} \\
\bottomrule
\end{tabular}
\end{table*}

\subsection{Experiment Setup}

\noindent
{\bf Dataset.}   
The memory fault prediction dataset comprises 9 months of memory logs. The first 5 months are designated as the training set and the subsequent 4 months as the test set. Since bit-level features can be represented as a $8 \times 4$ matrix, we compare several image classification methods in addition to existing memory failure prediction approaches. All algorithms were reproduced on our datasets to ensure a fair evaluation. Codes are publicly available at: \href{https://github.com/hwcloud-RAS/M2-MFP}{https://github.com/hwcloud-RAS/M2-MFP}.

\noindent
{\bf Baselines.}  
For the time-point approaches, only bit-level features are used since the DIMM-level information from an individual CE sample is insufficient for fault prediction. The \textbf{Naive} method compares bit-level feature frequencies in positive and negative samples, predicting a fault when the positive frequency is higher. \textbf{Risky CE}\cite{li2022correctable} predicts a fault when both lower bits DQ (positions 0 and 1) and higher bits DQ (positions 2 and 3) exhibit faults simultaneously. \textbf{DQ Beat Predictor}\cite{yu2023himfp} infers a fault when more than one faulty DQ and faulty Beat are present. \textbf{CNN}\cite{krizhevsky2012imagenet} represents bit-level features as a $8 \times 4$ matrix, applying an image classification approach. \textbf{Our Time-point module} is trained using a customized decision tree with a maximum depth of 4.


In the time-patch approaches, samples are generated using hourly aggregated data obtained by summing all bit-level features.  \textbf{CNN} uses a $3 \times 3$ convolution kernel; \textbf{CNN (1d kernel)} applies $1 \times 4$ and $8 \times 1$ kernels with max pooling. \textbf{ViT}\cite{dosovitskiy2020image} divides the $8 \times 4$ matrix into 32 patches of size $1 \times 1$; \textbf{ViT (1d patch)} uses $1 \times 4$ and $8 \times 1$ patches. \textbf{STIM}\cite{liustim} aggregates features over 6 consecutive hours and processes them with a Transformer encoder. All methods concatenate encoded bit-level features with CE count features and train an MLP classifier. \textbf{Himfp}\cite{yu2023himfp} integrates DIMM-level and bit-level count and pattern features, while we only utilize the publicly available features described in the paper to ensure reproducibility. Both Himfp and \textbf{our Time-patch module} generate high-level features using three observation windows (15 minutes, 1 hour, and 6 hours).


All deep learning methods are trained with a batch size of 256 using the Adam optimizer (learning rate 0.001) for 1000 epochs with early stopping (patience of 3). With the exception of the rule-based models, all models use cross-entropy as the loss function. See more details in Appendix \ref{sec:appendix-imple}. 


\noindent
{\bf Evaluation metrics.}  
Let \(t\) be the prediction timestamp, and define the prediction window as $
\mathcal{T}_{\text{pred}}(t) = \left[t+\Delta t_{\text{lead}},\, t+\Delta t_{\text{lead}}+\Delta t_\text{valid}\right]$, where \(\Delta t_{\text{lead}} = 15\) minutes (minimum lead time) and \(\Delta t_\text{valid} = 7\) days (prediction horizon). Let \(\mathcal{F}\) denote the set of DIMMs that experience failure during the test period, and let \(t_f^s\) denote the failure time of DIMM \(s\). A prediction for DIMM \(s\) at time \(t\) is considered correct if \(s \in \mathcal{F}\) and \(t_f^s \in \mathcal{T}_{\text{pred}}(t)\).  
Define the set of DIMMs that both experienced a failure and were successfully predicted by
\begin{equation}
    \mathcal{S} = \Bigl\{ s \in \mathcal{F} : \exists\, t \;\bigl( y_{\text{pred},s}(t)=1 \land t_f^s \in \mathcal{T}_{\text{pred}}(t) \bigr) \Bigr\}.
\end{equation}

Then, the evaluation metrics are computed as:
\begin{equation}
\text{P} = \frac{|\mathcal{S}|}{|{ s : \exists\, t \; y_{\text{pred},s}(t)=1 }|},\quad \text{R} = \frac{|\mathcal{S}|}{|\mathcal{F}|},\quad F_1 = \frac{2\times \text{P}\times\text{R}}{\text{P}+\text{R}}.
\end{equation}

More details of the evaluation can be found in Appendix \ref{sec:appendix-eval}.

\subsection{Performance Comparison}

We evaluate all baseline methods based on their best $F_1$-scores for a fair comparison. Meanwhile, \emph{Time-point-ours} (the Time-point module in M$^2$-MFP), \emph{Time-patch-ours} (the Time-patch module in M$^2$-MFP), and \emph{Combined-ours} (M$^2$-MFP) employ thresholds optimized through 5-fold cross-validation. As shown in Table \ref{tab:results_all}, our Time-point and Time-patch modules outperform their respective group counterparts, with further enhancement observed in the combined M$^2$-MFP framework, demonstrating approximately 55\% relative $F_1$-score improvement over the top baseline method. 

The results highlight limitations of rule-based methods (e.g., Risky CE, DQ Beat) in false positives, while adapted CNN (1D kernel) and ViT (2D patch) variants surpass standard versions, underscoring the importance of CE-specific temporal features. Crucially, 2D-BSF extracts CE patterns from bit-level information with three key advantages: symmetry (unlike naive methods), generality (unlike DQ-Beat/Himfp), and sensitivity (unlike CNN/ViT/STIM). By integrating these advancements into M$^2$-MFP's multi-scale hierarchy, we achieve state-of-the-art predictive performance.



\subsection{Ablation Experiments}

\subsubsection{Ablation Study on the Time-patch Module}

\begin{figure}[ht]
  \centering
  \includegraphics[width=\linewidth]{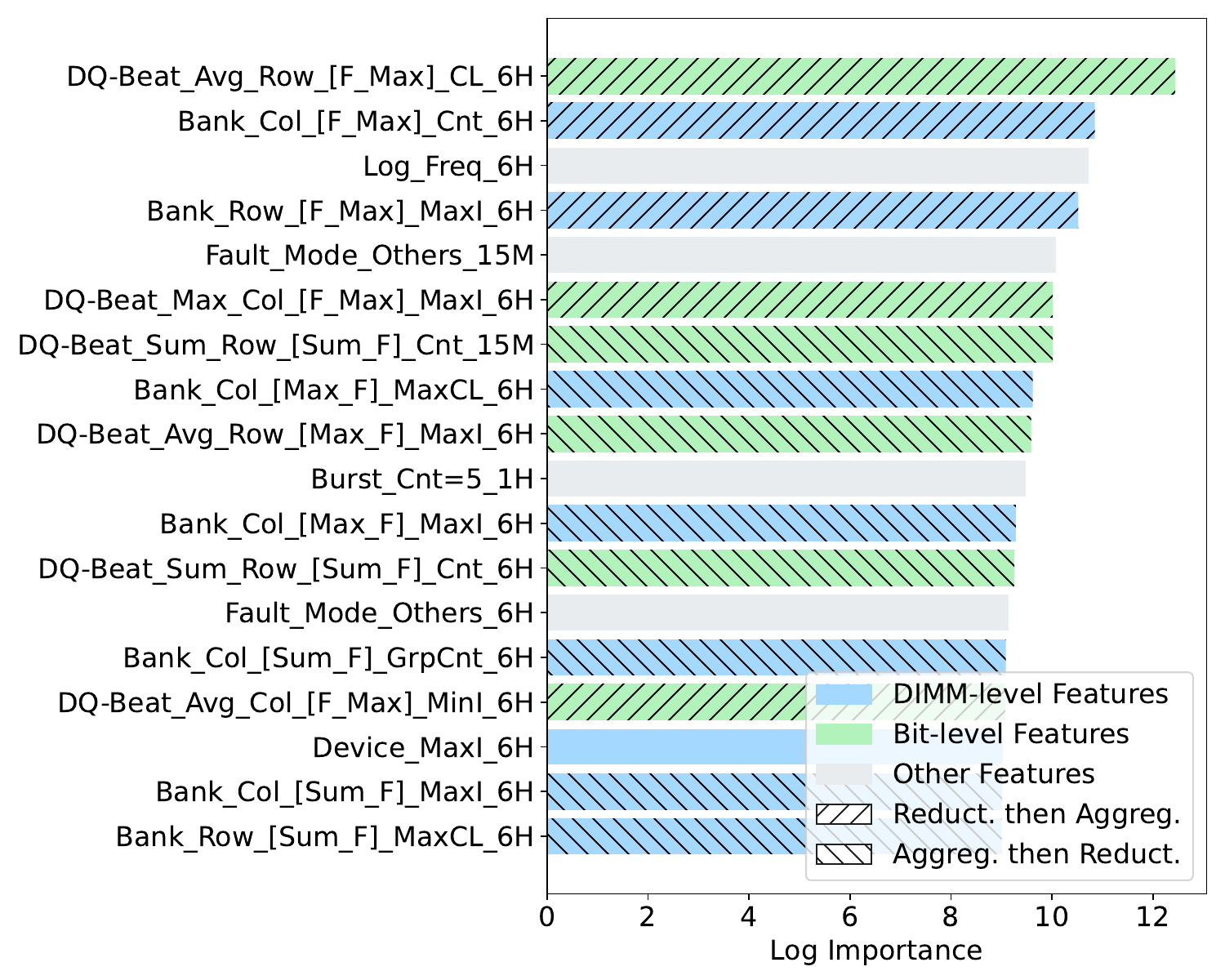}
  \caption{Feature Importance of the Time-Patch Module}
  \Description{time-patch-feature-importance}
  \label{fig:time-patch-feature-importance}
\end{figure}

We first analyze the feature importance in the LightGBM classifier used in the Time-patch module with multi-BSFE features. The evaluation of feature importance reveals critical patterns as shown in Figure~\ref{fig:time-patch-feature-importance}. It is evident that in the Time-patch module, both the DIMM-level features (illustrated by blue bars) and the bit-level features (depicted by green bars) extracted from the multi-BSFE are highly important. Additionally, features obtained through the Reduction-then-Aggregation (marked with left slashes “/”) and the Aggregation-then-Reduction (marked with right slashes “$\backslash$”) paths are identified as critical contributors to the model's performance.

We further conduct an ablation study comparing five configurations: (1) without the Reduction-then-Aggregation path, (2) without the Aggregation-then-Reduction path, (3) using only DIMM-level features, (4) using only bit-level features, and (5) our full approach. The experimental results (as shown in Table~\ref{tab:ablation_time_patch}) demonstrate the necessity of dual-path feature extraction and multi-level feature integration. 

\begin{table}[ht]
\captionsetup{skip=6pt}
\setlength{\tabcolsep}{7pt}
\centering
\caption{Time-patch Module Ablation Results}
\label{tab:ablation_time_patch}
\begin{tabular}{lccc}
\toprule
Method & Precision & Recall & $F_1$-score \\
\midrule
w/o Reduct. then Aggreg. & 0.3286 & 0.2725 & 0.2979 \\
w/o Aggreg. then Reduct. & 0.3563 & 0.2461 & 0.2911 \\
w/o DIMM-level features & 0.2980 & 0.2725 & 0.2847 \\
w/o bit-level features & 0.3744 & 0.1849 & 0.2475 \\
\midrule
\textbf{Time-patch Ours} & 0.3446 & 0.2893 & \textbf{0.3145} \\
\bottomrule
\end{tabular}
\end{table}

\subsubsection{Ablation Study on the Time-point Module}

For the Time-point Module ablation experiments, we maintain the same model input while replacing our designed Time-point Module with several alternative methods: LightGBM, XGBoost, FTTransformer, and a Gini decision tree. The experimental results (as shown in Table~\ref{tab:ablation_time_point}) clearly indicate that our proposed Time-point Module achieves the best performance compared to these alternatives.

\begin{table}[ht]
\captionsetup{skip=6pt}
\setlength{\tabcolsep}{7pt}
\centering
\caption{Time-point Module Ablation Results}
\label{tab:ablation_time_point}
\begin{tabular}{lcccc}
\toprule
Method & Precision & Recall & $F_1$-score \\
\midrule
LightGBM & 0.0916 & 0.3349 & 0.1439 \\
XGBoost & 0.0951 & 0.3373 & 0.1484 \\
FTtransformer & 0.0664 & 0.2749 & 0.1069 \\
Decision Tree (Gini) & 0.1036 & 0.3433 & 0.1591 \\
\midrule
\textbf{Time-point Ours} & 0.4029 & 0.2245 & \textbf{0.2883} \\
\bottomrule
\vspace{-0.5cm}
\end{tabular}
\end{table}

\begin{figure}[ht]
  \centering
  \includegraphics[width=\linewidth]{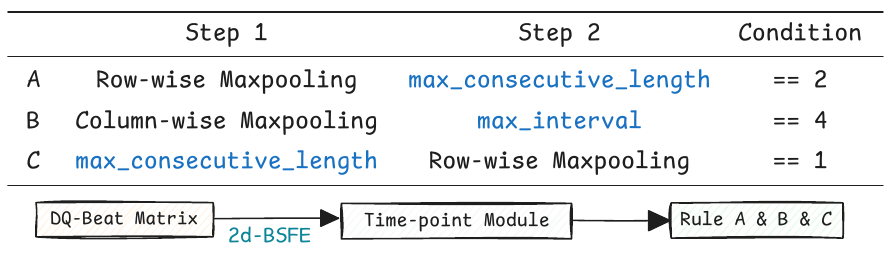}
  \caption{A rule identified by the Time-point Module}
  \Description{fault-pattern}
  \label{fig:fault_pattern}
\end{figure}

A rule learned from 2d-BSFE and identified as a fault pattern by the Time-point Module is A \& B \& C, where A and B are Aggregation-then-Reduction conditions, and C is Reduction-then-Aggregation condition, as shown in Figure \ref{fig:fault_pattern}.

\subsection{Lead Time Analysis}

\begin{figure}[ht]
  \centering
  \includegraphics[width=0.9\linewidth]{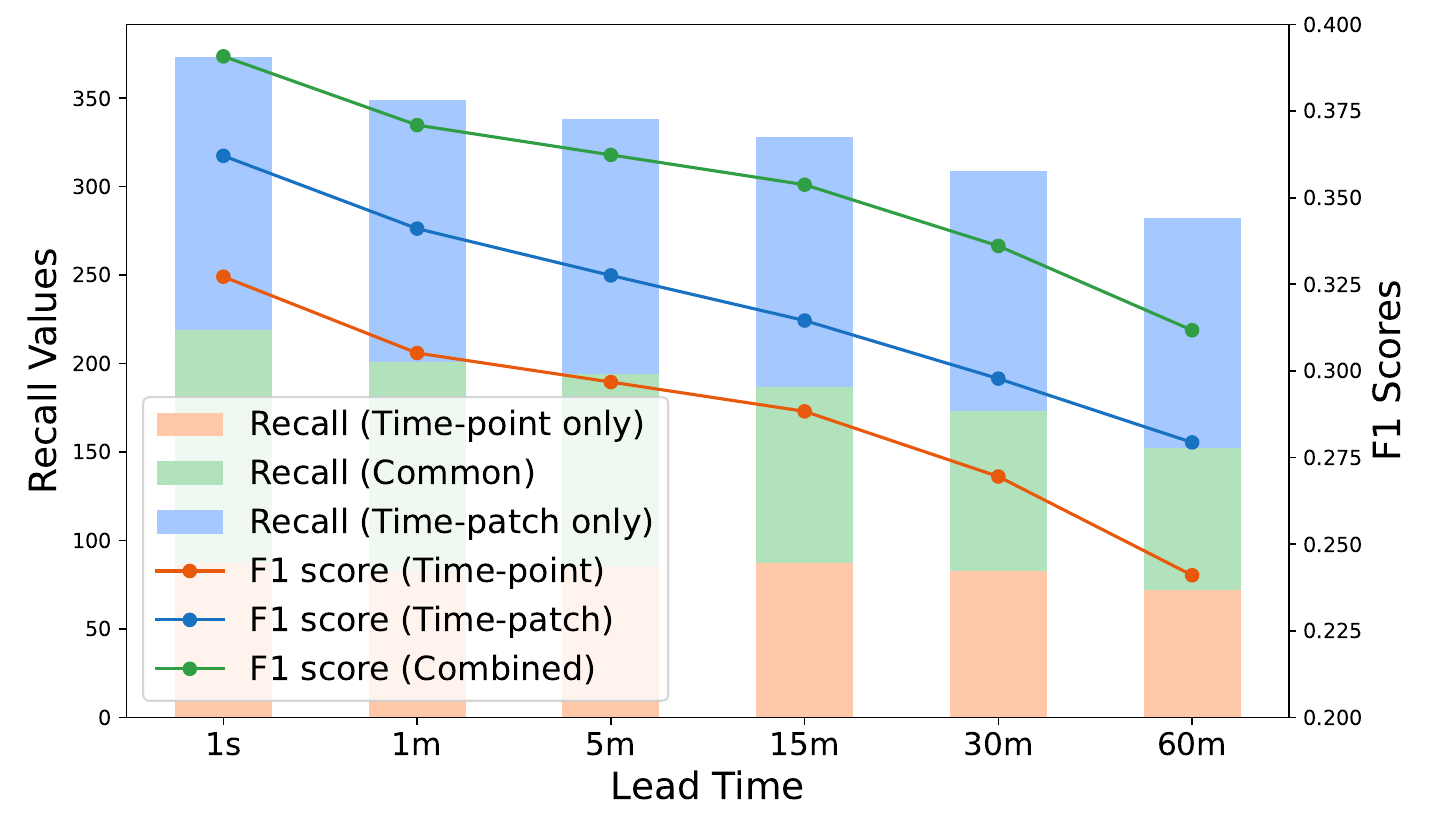}
  \caption{Variation of Recall and F1 Score Across Different Lead Times}
  \Description{lead-time-analysis}
  \label{fig:lead-time-analysis}
\end{figure}

In practical applications, due to data transmission delays and the time required for failure handling (e.g., failure isolation or hot migration), predictive algorithms must anticipate failures in advance. We analyzed the impact of lead time by varying $\Delta t_{\text{lead}}$  from 1 second to 60 minutes. The performance of M$^2$-MFP and its variants is illustrated in Figure \ref{fig:lead-time-analysis}. The analysis shows that as $\Delta t_{\text{lead}}$ increases, the number of recalled failed DIMMs decreases. However, the two modules of Multi-scale MFP remain complementary across all settings of $\Delta t_{\text{lead}}$, and their combined use effectively improves both the failure recall rate and the $F_1$-score.

\section{Deployment}

\subsection{Deployment Details}
\label{sec:deployment}
The Huawei Cloud Memory Failure Prediction System operates through four key stages, following a workflow similar to that shown in Figure \ref{fig:AIOps-System}. First, server hardware generates raw error signals via embedded sensors and memory controllers. These signals are processed by the Baseboard Management Controller (BMC), which monitors equipment health in real-time, converts sensor data into timestamped CE logs, and transmits the events to upstream systems. Next, the data flows to the AIOps System, where CE events are collected in real time, complete CE sequences are archived in a data warehouse, and predictive analysis is performed using the M$^2$-MFP framework. When the risk score exceeds a predefined threshold, an alert mechanism is activated. Finally, prediction alerts are transmitted to the Cloud Monitor Alarm (CMA) platform, which coordinates maintenance operations through the enterprise IT infrastructure.

Upon predicting a memory failure, the SRE operations system promptly initiates live service migration and server decommissioning. Subsequently, stress testing is performed on the memory modules. If memory issues are confirmed, the faulty DIMMs are replaced before services are migrated back. It should be noted that memory recovery techniques are not universally applicable—each server adopts appropriate remediation measures based on its usage scenario and RAS (Reliability, Availability, Serviceability) objectives.

\subsection{Online Validation}
\label{sec:online_validation}
Validation was conducted in Huawei Cloud's reliability gray environment using Intel Purley platform data (January–August 2024). The dataset was partitioned into 5-month training and 3-month test phases. Compared to UniMFP, the enhanced version of Himfp that deployed in the live production, M$^2$-MFP demonstrates a 15\% improvement in $F_1$-score and over a 20\% precision gain at high-recall regimes (as shown in Table~\ref{tab:online-results}). It is important to note that the differences between the online and offline evaluation results can be largely attributed to the data volume and composition: the online evaluation benefits from a larger dataset with a substantially higher number of positive samples, thereby providing more comprehensive failure pattern information that enhances model performance. 
M$^2$-MFP achieves better results without relying on domain-specific expert knowledge, thereby demonstrating the robustness and generalizability of our universal framework in diverse operational settings.

In our production environment, the system is configured with a 15-minute inference interval and Multi-BSFE aggregation across three temporal scales. The average end-to-end inference time is approximately 4 minutes, with a peak time of around 7 minutes, which is entirely acceptable in practice.Currently, M$^2$-MFP is deployed in the production environment, providing continuous and reliable failure prediction for millions of memory devices.

\begin{table}[ht]
\setlength{\tabcolsep}{3pt}
\captionsetup{skip=6pt}
\centering
\caption{Online validation results (values sanitized to 2 significant digits)}
\label{tab:online-results}
\begin{tabular}{cccccc}
\toprule
Method & Precision & Recall & $F_1$-score & P@R=0.5 & P@R=0.6 \\
\midrule
UniMFP & 0.29 & 0.38 & 0.33 & 0.23 & 0.18 \\
M\textsuperscript{2}-MFP & \textbf{0.32} & \textbf{0.45} & \textbf{0.38} & \textbf{0.29} & \textbf{0.22} \\
\midrule
Rel. Improv.  & 10\% & 18\% & 15\% & 26\% & 22\% \\
\bottomrule
\end{tabular}
\end{table}

\section{Conclusion}

In this work, we present M$^2$-MFP, a novel multi-scale and multi-level framework for proactive memory failure prediction. M$^2$-MFP integrates a multi-level binary spatial feature extractor with dual-path temporal modeling to effectively capture critical patterns from Correctable Errors (CEs). Our approach overcomes key challenges such as operational noise, data missingness, extreme class imbalance, and hardware heterogeneity, as demonstrated by comprehensive evaluations on a large-scale real-world dataset from Huawei Cloud. The notable performance enhancements, a roughly 15\% boost in the $F_1$-score, and its successful deployment on Huawei Cloud's AIOps platform highlight the practical value of M$^2$-MFP in guaranteeing the stability of cloud infrastructure operations and providing insights for more intelligent and automated proactive maintenance systems.

\begin{acks}

The work was supported by grants from the National Natural Science Foundation of China (No. 92367110). Thanks to Huawei Cloud for data and resources supporting.

\end{acks}

\bibliographystyle{ACM-Reference-Format}

\balance

\bibliography{references}

\appendix

\section*{Appendix}

\section{Dataset Details}
\label{sec:appendix-data}


Our dataset comprises log data from \emph{Intel Purley} and \emph{Intel Whitley} DIMMs, collected from over 70,000 DIMMs between January and September 2024. For each month, we computed the number of DIMMs containing CE logs, the number of DIMMs that experienced failures, and the total number of CE events recorded. The results are presented in Table~\ref{tab:summary}. Note that the sum of the monthly DIMM counts does not match the total DIMM count because some DIMMs exhibit CE logs in multiple months.

Table~\ref{tab:celog} lists the fields contained in each CE log, which are collected via mcelog into 23 columns. Note that sensitive information, including manufacturer and region, has been anonymized in accordance with our confidentiality policy.

\begin{table*}[ht]
\centering
\caption{Summary of DIMM and CE counts}
\label{tab:summary}
\begin{tabular}{l *{3}{c} *{3}{c} *{3}{c}}
\toprule
& \multicolumn{3}{c}{Intel Purley} & \multicolumn{3}{c}{Intel Whitley} & \multicolumn{3}{c}{all} \\
\cmidrule(lr){2-4} \cmidrule(lr){5-7} \cmidrule(lr){8-10}
Month & DIMM & Fault DIMM & CE & DIMM & Fault DIMM & CE & DIMM & Fault DIMM & CE \\
& count & count & count & count & count & count & count & count & count \\
\midrule
2024-01 & 26450 & 163 & 102423615 & 1466 & 15 & 1843056 & 27916 & 178 & 104266671 \\
2024-02 & 24682 & 133 & 97336381 & 1682 & 8 & 1763660 & 26364 & 141 & 99100041 \\
2024-03 & 26919 & 174 & 106745981 & 1948 & 16 & 1682764 & 28867 & 190 & 108428745 \\
2024-04 & 29776 & 144 & 107168332 & 2067 & 17 & 1947564 & 31843 & 161 & 109115896 \\
2024-05 & 29046 & 155 & 119407409 & 2191 & 12 & 2494964 & 31237 & 167 & 121902373 \\
2024-06 & 32610 & 169 & 118064609 & 2270 & 16 & 2580721 & 34880 & 185 & 120645330 \\
2024-07 & 37853 & 250 & 148794345 & 2564 & 19 & 3604404 & 40417 & 269 & 152398749 \\
2024-08 & 37531 & 214 & 165111964 & 2631 & 18 & 3564878 & 40162 & 232 & 168676842 \\
2024-09 & 37613 & 160 & 158444298 & 2668 & 21 & 3296809 & 40281 & 181 & 161741107 \\
\midrule
Total & 64794 & 1562 & 1123496934 & 7175 & 142 & 22778820 & 71969 & 1704 & 1146275754 \\
\bottomrule
\end{tabular}
\end{table*}

\begin{table*}[ht]
  \centering
  \caption{DRAM CE Log Explanation}
  \begin{tabular}{rllll}
    \toprule
    No. & Field & Level & Type & Description \\
    \midrule
    1  & cpuid              & server & integer  & The CPU ID, note that a server attaches multiple CPUs. \\
    2  & channelid          & server & integer  & The Channel ID, note that a CPU has multiple channels. \\
    3  & dimmid             & server & integer  & The DIMM ID, note that a channel attaches multiple DIMMs. \\
    4  & rankid             & DIMM & integer  & The rank ID, range from 0 to 1, each DIMM has 1 or 2 ranks. \\
    5  & deviceid           & DIMM & integer  & The device ID, range from 0 to 17, each DIMM has multiple devices. \\
    6  & bankgroupid        & DIMM & integer  & The bank group ID of DRAM. \\
    7  & bankid             & DIMM & integer  & The bank ID of DRAM. \\
    8  & rowid              & DIMM & integer  & The row ID of DRAM. \\
    9  & columnid           & DIMM & integer  & The column ID of DRAM. \\
    10 & retryrderrlogparity& bit & integer  & The parity info on retried reads in decimal format. \\
    11 & retryrderrlog      & bit & integer  & Logs info on retried reads in decimal format, validating error types and parity. \\
    12 & beat\_info        & bit & integer  & The decoded parity error bits in memory DQ and Beat. \\
    13 & error\_type        & other & integer  & The error type including read and scrubbing error. \\
    14 & log\_time          & other & string   & The time when the error is detected in timestamp. \\
    15 & manufacturter      & server & category & The server manufacturer in anonymized format. \\
    16 & model              & server & category & The CPU model in anonymized format. \\
    17 & PN                 & server & category & The part number of DIMMs in anonymized format. \\
    18 & Capacity           & server & integer  & The capacity of DIMM. \\
    19 & FrequencyMHz       & server & integer  & Base frequency of the CPU resource, in MHz. \\
    20 & MaxSpeedMHz        & server & integer  & Maximum frequency of the CPU resource. \\
    21 & McaBank            & server & category & Machine Check Architecture bank code of the CPU. \\
    22 & memory\_type       & server & string   & The type of DIMM, e.g., DDR4. \\
    23 & region             & server & category & The region of server location in anonymized format. \\
    \bottomrule
  \end{tabular}
  \label{tab:celog}
\end{table*}

\section{Implementation Details}
\label{sec:appendix-imple}

\subsection{Time-point Methods}

\subsubsection{Naive Method}
Each CE is represented by a DQ-Beat Matrix $B(e) \in \{0,1\}^{8 \times 4}$, where rows are beats and columns are DQs. Let $n_{+}(B(e))$ and $n_{-}(B(e))$ count occurrences in positive and negative samples. 

A CE is flagged as faulty if:
\[ n_{+}(B(e)) > n_{-}(B(e)). \]

\subsubsection{Risky CE Rule}
This rule examines bit-level errors in $B(e)$ across DQ signals $\text{DQ}_0$ to $\text{DQ}_3$. 

A CE is faulty if:
\[ (e(\text{DQ}_0) + e(\text{DQ}_1) \geq 1) \land (e(\text{DQ}_2) + e(\text{DQ}_3) \geq 1). \]

\subsubsection{DQ Beat Predictor}
Counts errors in DQs and Beats:
\[ N_{\text{DQ}} = \text{DQs with errors}, \quad N_{\text{Beat}} = \text{Beats with errors}. \]
A CE is faulty if:
\[ N_{\text{DQ}} > 1 \land N_{\text{Beat}} > 1. \]

\subsubsection{CNN-based Method (Time-point)}
In this approach, the input is the binary DQ-Beat Matrix
\[
\mathbf{X} \in \{0,1\}^{8 \times 4},
\]
representing the CE's bit-level feature (equals to B(e)). This method adopts a hybrid CNN-MLP architecture similar to that used in the time-patch methods. Specifically, the model processes inputs via parallel pathways:
\begin{align}
    \mathbf{H}_{\text{CNN}} &= \sigma\Big(\mathbf{W}_{\text{conv}} \ast \mathbf{X} + \mathbf{b}_{\text{conv}}\Big), \\
    \mathbf{H}_{\text{MLP}} &= \sigma\Big(\mathbf{W}_2 \cdot \sigma\big(\mathbf{W}_1 \mathbf{x} + \mathbf{b}_1\big) + \mathbf{b}_2\Big), \\
    \mathbf{y} &= \text{Softmax}\Big(\mathbf{W}_c \Big[\mathbf{H}_{\text{CNN}} \oplus \mathbf{H}_{\text{MLP}}\Big] + \mathbf{b}_c\Big),
\end{align}
where \(\ast\) denotes the convolution operation, \(\oplus\) denotes concatenation, and \(\sigma\) is the ReLU activation function. Here, \(\mathbf{x} \in \mathbb{R}^{8}\) represents the DIMM-level CE statistical features.

\subsection{Time-patch Methods}

\subsubsection{CNN-based Method}
The model combines a CNN for image processing and an MLP for statistical features:

\begin{itemize}[leftmargin=*]
    \item \textbf{CNN Branch:}  
    - Input: 8×4 DQ-Beat Matrix  
    - Layers: 3×3 conv (32 filters, padding 1) → 2×2 max pool → flatten → FC(128)  
    - Regularization: Dropout (p=0.1) after FC layer
    
    \item \textbf{MLP Branch:}  
    - Input: 8-dim CE features  
    - Layers: FC(8→64) → Dropout (0.1) → FC(64→128)
\end{itemize}

The combined feature $\mathbf{h}_{\text{combined}} \in \mathbb{R}^{256}$ is classified via:
\[ \mathbf{y} = \mathbf{W}_{\text{combined}}\mathbf{h}_{\text{combined}}, \quad \mathbf{W} \in \mathbb{R}^{2 \times 256} \]

\subsubsection{1D CNN-based Method}
Dual 1D convolutional pathways:

\begin{itemize}[leftmargin=*]
    \item \textbf{Row Pathway:}  
    (1,4) conv → (8,1) max pool (stride 8,1) → additional convs → 32 features
    
    \item \textbf{Column Pathway:}  
    (8,1) conv → (1,4) max pool (stride 1,4) → additional convs → 32 features
\end{itemize}

Features are concatenated (128D) → FC(128) → dropout (0.1). Combined with MLP branch (same as above) for final classification.

\subsubsection{ViT-based Method}
Processes image patches ($p \times p$) via transformer:

\begin{itemize}[leftmargin=*]
    \item Patch embedding: 1×1 patches ($p=1$) → 128D embedding
    \item Class token + learned positional embeddings
    \item Transformer: 3 layers, 4 heads, 256-dim MLP
    \item Final classification uses class token representation
\end{itemize}

MLP branch processes 8-dim features separately. Combined features classified via linear layer.

\subsubsection{1D ViT-based Method}
Dual-branch 1D ViT:

\begin{itemize}[leftmargin=*]
    \item \textbf{Branch 1:} 8-length patches (N=4), 2 layers
    \item \textbf{Branch 2:} 4-length patches (N=8), 2 layers 
    \item Both use 4 attention heads, d=128
    \item Combines with MLP branch (d=128)
\end{itemize}

Final classification on $\mathbb{R}^{384}$ combined features.

\section{Evaluation Details}
\label{sec:appendix-eval}

Memory Failure Prediction models must meet the following requirements to accommodate memory failure prediction tasks based on streaming data in real-world scenarios:

\begin{itemize}[leftmargin=*]
    \item When predicting whether a failure will occur in the future at time \(t\), the model may only use log data available before time \(t\);
    \item For each prediction, the assigned prediction timestamp must not be earlier than the timestamp of the latest log entry in the input data;
    \item Historical predictions must remain unaltered by future predictions—i.e., a prediction made at time \(t\) should not be affected by predictions made after \(t\).
\end{itemize}

Let \(t\) denote the prediction timestamp and define the prediction window as
\[
\mathcal{T}_{\text{pred}}(t) = \Bigl[t + \Delta t_{\text{lead}},\; t + \Delta t_{\text{lead}} + \Delta t_p\Bigr],
\]
where \(\Delta t_{\text{lead}} = 15\) minutes is the minimum lead time and \(\Delta t_p = 7\) days extends the prediction horizon. Let \(\mathcal{F}\) denote the set of DIMMs that experience failure during the test period, and let \(t_f^s\) denote the failure time of DIMM \(s\).

A prediction for DIMM \(s\) at time \(t\) is considered correct if
\[
y_{\text{pred},s}(t) = 1 \quad \text{and} \quad t_f^s \in \mathcal{T}_{\text{pred}}(t).
\]

We define the following sets to formalize the evaluation metrics:

\begin{enumerate}
    \item \textbf{Predicted DIMMs:} The set of all DIMMs for which at least one failure prediction was issued during the test period:
    \begin{equation}
    \mathcal{S}_{\text{pred}} = \Bigl\{ s \,\Bigl|\, \exists\, t \text{ (in the test period) such that } y_{\text{pred},s}(t)=1 \Bigr\}.
    \end{equation}

    \item \textbf{Correctly Predicted DIMMs:} The set of DIMMs that (i) actually fail during the test period, and (ii) have at least one prediction issued at time \(t\) with the failure time \(t_f^s\) falling within the corresponding prediction window:
    \begin{equation}
    \mathcal{S}_{\text{true}} = \Bigl\{ s \in \mathcal{F} \,\Bigl|\, \exists\, t \text{ such that } y_{\text{pred},s}(t)=1 \text{ and } t_f^s \in \mathcal{T}_{\text{pred}}(t) \Bigr\}.
    \end{equation}
\end{enumerate}

The evaluation metrics are then calculated as follows:

\textbf{Precision:} This metric measures the proportion of correctly predicted DIMMs among all DIMMs that received at least one prediction. Its numerator and denominator are given by:
\begin{align}
    \text{Precision Numerator} &= |\mathcal{S}_{\text{true}}|, \\
    \text{Precision Denominator} &= |\mathcal{S}_{\text{pred}}|,
\end{align}
and thus,
\begin{equation}
\text{Precision} = \frac{|\mathcal{S}_{\text{true}}|}{|\mathcal{S}_{\text{pred}}|}.
\end{equation}

\textbf{Recall:} This metric quantifies the proportion of actual failures that were correctly predicted. Its numerator is the same as that of precision, and the denominator is the total number of DIMMs that failed during the test period:
\begin{align}
    \text{Recall Numerator} &= |\mathcal{S}_{\text{true}}|, \\
    \text{Recall Denominator} &= |\mathcal{F}|,
\end{align}
leading to:
\begin{equation}
\text{Recall} = \frac{|\mathcal{S}_{\text{true}}|}{|\mathcal{F}|}.
\end{equation}

\textbf{F\textsubscript{1} Score:} Finally, the F\textsubscript{1} score, which is the harmonic mean of precision and recall, is defined as:
\begin{equation}
F_1 = \frac{2 \cdot \text{Precision} \cdot \text{Recall}}{\text{Precision} + \text{Recall}}.
\end{equation}

Here, the operator \(|\cdot|\) denotes the cardinality of a set.


\end{document}